\documentclass[10pt,letterpaper]{article}
\usepackage[top=0.85in,left=2.75in,footskip=0.75in,marginparwidth=2in]{geometry}

\usepackage[utf8]{inputenc}

\usepackage{cite}

\usepackage{nameref,hyperref}

\usepackage[right]{lineno}

\usepackage{microtype}
\DisableLigatures[f]{encoding = *, family = * }

\raggedright
\usepackage{changepage}

\usepackage[aboveskip=1pt,labelfont=bf,labelsep=period,singlelinecheck=off]{caption}

\makeatletter
\renewcommand{\@biblabel}[1]{\quad#1.}
\makeatother

\usepackage{lastpage,fancyhdr,graphicx}
\usepackage{epstopdf}
\fancyheadoffset[L]{2.25in}
\fancyfootoffset[L]{2.25in}

\usepackage{color}

\definecolor{Gray}{gray}{.25}

\usepackage{graphicx}

\usepackage{sidecap}

\usepackage{wrapfig}
\usepackage[pscoord]{eso-pic}
\usepackage[fulladjust]{marginnote}
\reversemarginpar

\begin{document}
\vspace*{0.35in}

\begin{flushleft}
{\Large
\textbf\newline{Vibrational and electronic properties of Al$_{12}$M (M=Cu, Zn) clusters: DFT calculations}
}
\newline
\\
P. L. Rodr\'iguez-Kessler\textsuperscript{1,*}
\\
\bigskip
\bf{1} Centro de Investigaciones en \'Optica A.C., Loma del Bosque 115, Lomas del Campestre , Leon, 37150, Guanajuato, Mexico
\bigskip
*plkessler@cio.mx

\end{flushleft}

\justifying

\section*{Abstract}
 In this work, the vibrational and electronic properties of Al$_{12}$M (Zn, Cu) clusters are investigated using density functional theory (DFT) calculations. The results indicate that the clusters favor low-spin states when evaluated with three different functionals: PBE, PBE0, and TPSSh. Additionally, the doped clusters exhibit lower ionization energy and electron affinity compared to the neutral Al$_{13}$ cluster. The density of states show a higher degree of hybridization in Al$_{12}$Cu compared to Al$_{12}$Zn.

\section*{Introduction}

Aluminum clusters and nanoparticles have garnered significant interest from researchers due to their wide range of potential applications. In particular, aluminium-based nanoclusters have shown promising properties for clean energy applications.\cite{GRAETZ2011S517,ZUTTEL20031,BOGDANOVIC2007813} The growth mechanism, electronic properties and spectra of aluminum clusters with 3–20 atoms has been studied by Tan et al.\cite{TAN2022120545} They revealed that the Al$_{7}^+$ and Al$_{13}^-$ clusters exhibit very high stability and a large energy gap, and can be regarded as superatoms. Moreover, the absorption spectra of neutral aluminum clusters have been simulated and compared with experimental findings.\cite{PhysRevB.60.R11297} The Al$_{13}$ cluster, in particular, is widely studied due to its superatom characteristics in the anionic state. Additionally, the icosahedral structure of Al$_{13}$ makes it an attractive system with a well-defined geometry.\cite{10.1063/1.3075834} In order to enhance the catalytic properties through specific applications, a study of both exohedral and endohedral doping on the Al$_{13}$ cluster has been conducted. For example, endohedral and exohedral magnesium-doped aluminum clusters were evaluated for their hydrogen adsorption properties.\cite{doi:10.1021/jp911013t} Vanbuel et al. evaluated the H$_2$ chemisorption on Al$_{n}$Rh$^{2+}$ (n=10-13) clusters and found that for n=10 and 11, a single H$_2$ molecule binds dissociatively, whereas for n=12 and 13, it adsorbs molecularly.\cite{Vanbuel2018} Although many studies have evaluated the hydrogen adsorption properties of transition metal-doped Al clusters, only a few have examined their catalytic activities, and further progress in this area is still needed.\cite{Das2014} Recently, Samudre et al. investigated the endohedral doping of X atoms (X = Ti, V, Fe, Co, Ni, Cu, Zn, Y, Mo, Ru, Rh, and W) in the Al$_{13}$ cluster and revealed that only Al$_{12}$Zn and Al$_{12}$Cu are thermally stable endohedral clusters, according to BOMD simulation results.\cite{doi:10.1080/08927022.2022.2153151} However, further descriptive parameters of the clusters have not been evaluated so far. In this brief preprint, we investigate the vibrational and electronic properties of endohedral Al$_{12}$Zn and Al$_{12}$Cu clusters. The doped clusters show lower ionization energy and electron affinity compared to the neutral Al$_{13}$ cluster. The density of states show a higher degree of hybridization in Al$_{12}$Cu compared to Al$_{12}$Zn.

\section*{Materials and Methods}

The calculations in this study utilize density functional theory (DFT) as implemented in the Orca quantum chemistry package. \cite{10.1063/5.0004608}. The Exchange and correlation energies are addressed using various functionals, including the GGA (PBE),\cite{PhysRevLett.77.3865} meta-GGA (TPSS)\cite{PhysRevLett.91.146401} and hybrid (PBE0)\cite{10.1063/1.478522} in conjunction with the Def2-TZVP basis set.\cite{B508541A} Atomic positions are self-consistently relaxed through a Quasi-Newton method employing the BFGS algorithm. The SCF convergence criteria for geometry optimizations are achieved when the total energy difference is smaller than 10$^{-8}$ au, by using the TightSCF keyword in the input. The  Van  der  Waals  interactions  are  included in the exchange-correlation functionals with empirical dispersion corrections of Grimme DFT-D3(BJ). The total density of states (DOS) and partial density of states (PDOS) for clusters and complexes were obtained by using the Multiwfn program.\cite{https://doi.org/10.1002/jcc.22885}


\section*{Results}

\begin{figure}[h!]
  \vspace{.5cm}
\begin{tabular}{p{2.0cm}p{2.0cm}p{2.0cm}}
\resizebox*{0.20\textwidth}{!}{\includegraphics{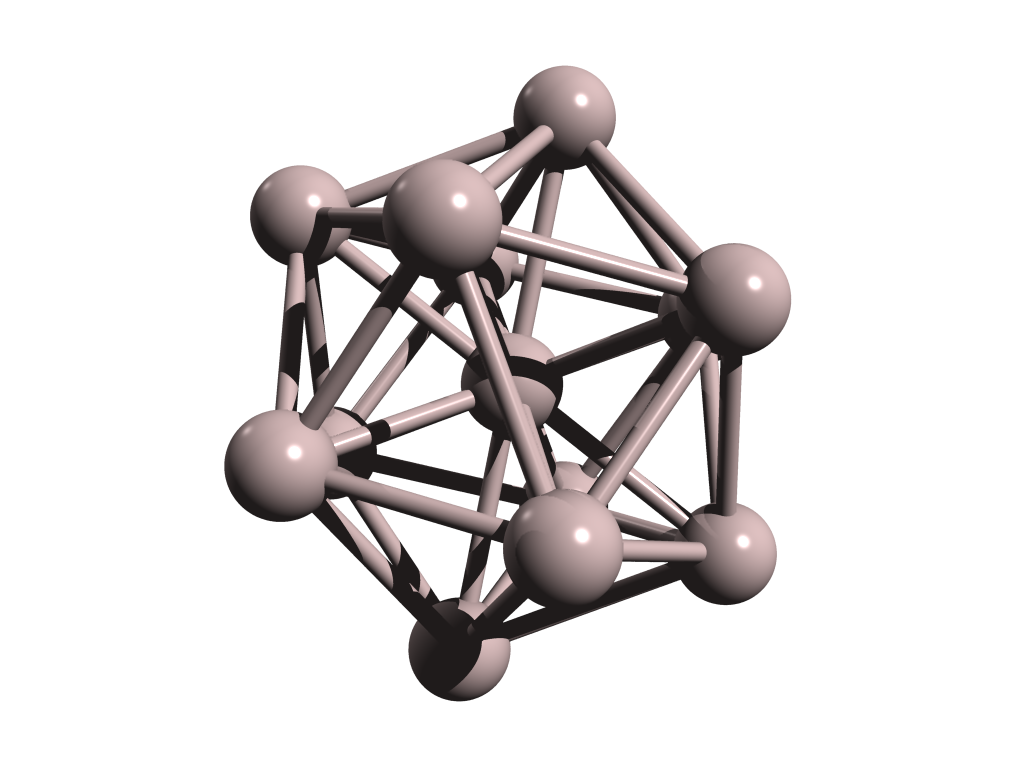}} &
\resizebox*{0.20\textwidth}{!}{\includegraphics{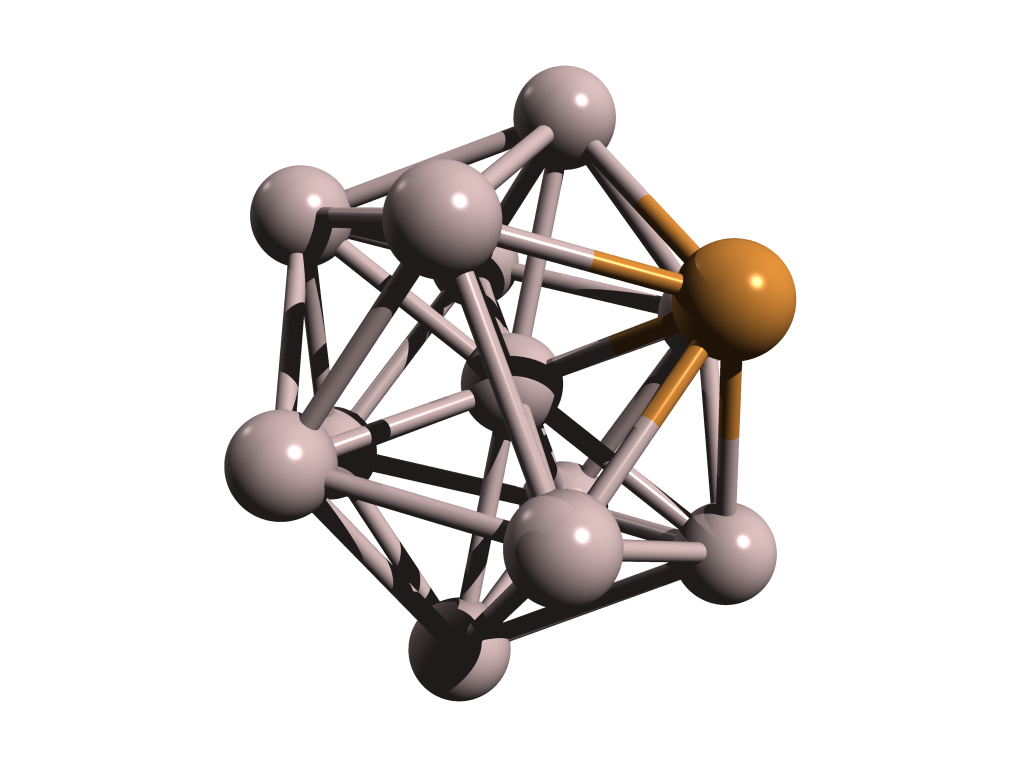}} &
\resizebox*{0.20\textwidth}{!}{\includegraphics{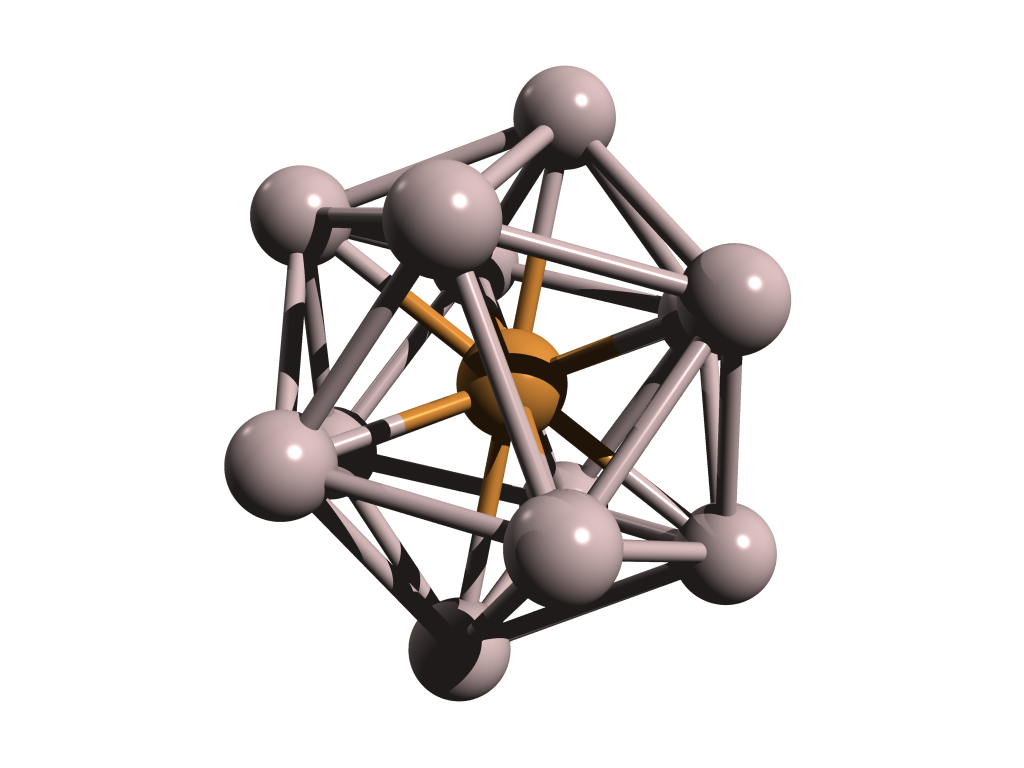}}   \\
  \hspace{0.8cm}   {\bf Pure}  &
  \hspace{0.2cm}   {\bf M-surface}  &
  \hspace{0.3cm}   {\bf M-center}  \\
       \end{tabular}
       \caption{\label{figure1}Representative icosahedral structures for Al$_{13}$ and Al$_{12}$M clusters.}
\end{figure}

 The structure models for representing the Al$_{12}$M (M=Cu, Zn) clusters are shown in Figure~\ref{figure1}. The results show that the clusters favor the endohedral doping, however, we found that for Al$_{12}$Cu the low-spin state (doublet) is favored over the quartet state, which is in contrast with the report of Samudre et al.\cite{doi:10.1080/08927022.2022.2153151} We have further confirmed the results by using three different functionals (Table~\ref{table1}). The PBE and PBE0 functionals favor the doublet and singlet states for Al$_{12}$Cu and Al$_{12}$Zn clusters, while the TPSSh functional favor the doublet and singlet states by 0.001 eV and 0.04 eV of the total energy for the Al$_{12}$Cu and Al$_{12}$Zn clusters, respectively.

\begin{table}[h!]
\caption{\label{table1}Relative energies (in eV) of the lowest energy structures of Al$_{12}$M (Cu, Zn) clusters computed at different DFT levels. Representative values on the spin multiplicity (S$_M$) are given.}
\begin{tabular}{p{1.8cm}p{0.8cm}p{0.8cm}p{0.8cm}p{0.8cm}}
{Label}  & S$_M$ & \small{PBE} & \small{PBE0} & \small{TPSSh} \\
\hline
\textbf{Al$_{12}$Zn} & 1 & 0.00 & 0.00 & 0.00 \\
\textbf{Al$_{12}$Cu}$^{a}$ & 2 & 0.00 & 0.00 & 0.00 \\
\\
\textbf{Al$_{12}$Zn} & 3 & 0.10 & 0.05 & 0.04 \\
\textbf{Al$_{12}$Cu}$^{b}$ & 4 & 0.17 & 0.03 & 0.00 \\
\hline
\textsuperscript{a}This work.\\
\textsuperscript{b}Ref.\citenum{doi:10.1080/08927022.2022.2153151}
\end{tabular}
\end{table}

\begin{figure}[h!]
\begin{center}
\scriptsize
 \includegraphics[scale=0.45]{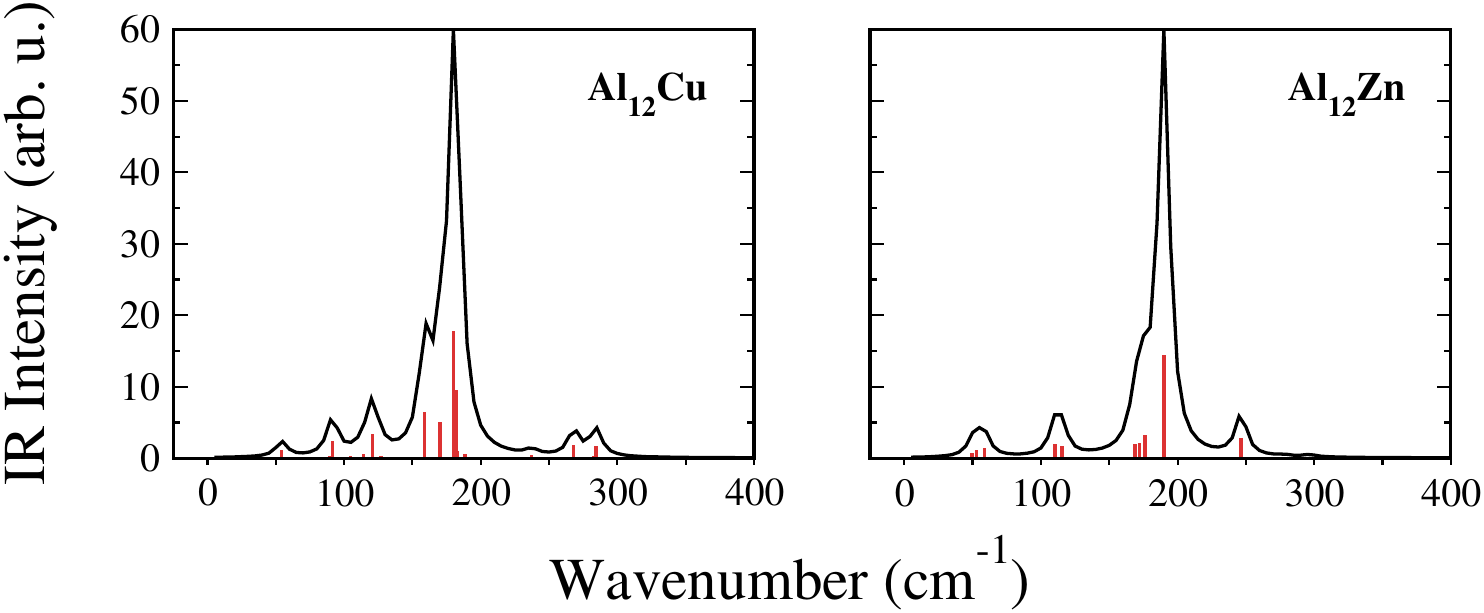}
	\caption{\label{figure_IR}IR spectra for Al$_{12}$Cu and Al$_{12}$Zn clusters obtained at the PBE0/Def2-TZVP level.}
\end{center}
\end{figure}

\begin{figure}[h!]
\Large
  \vspace{.5cm}
\begin{tabular}{cccccc}
\resizebox*{0.15\textwidth}{!}{\includegraphics{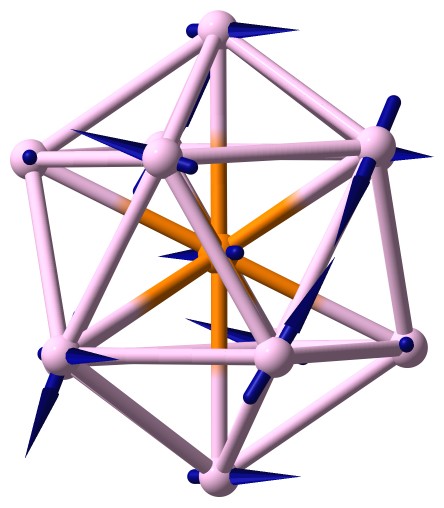}} &
\resizebox*{0.15\textwidth}{!}{\includegraphics{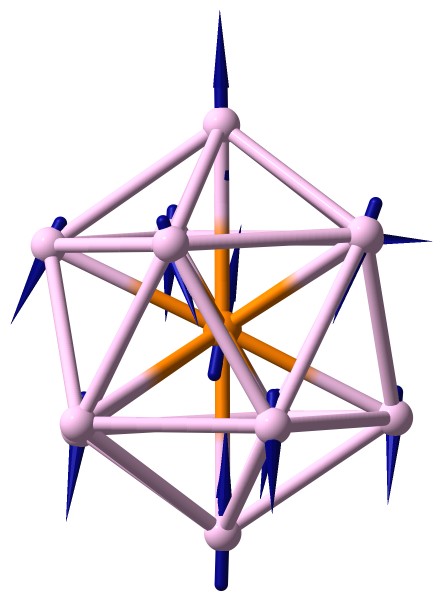}} &
\resizebox*{0.15\textwidth}{!}{\includegraphics{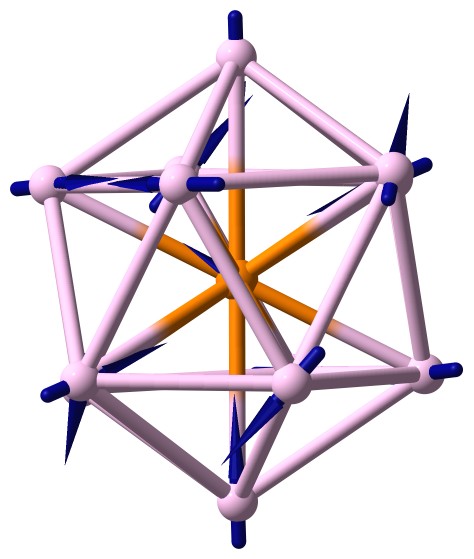}} &
   \\
   \small 54.22 cm$^{-1}$  &
   \small 180.02 cm$^{-1}$  &
   \small 305.41 cm$^{-1}$  &
       \end{tabular}
       \caption{\label{figure2}Representative vibrational modes for the Al$_{12}$Cu cluster.}
\end{figure}

To identify the fingerprints of the Al$_{12}$Cu and Al$_{12}$Zn clusters, we have calculated the infrared (IR) spectra, which serve as a guide for future experimental studies when become available.\cite{2024structuresstabilitiesb7cr2clusters,2024revisitingglobalminimumstructure} The characteristic peak for Al$_{12}$Cu is found at 180.02 cm$^{-1}$, while for Al$_{12}$Zn is found at 190.14 cm$^{-1}$ (Figure~\ref{figure_IR}). The lowest and highest vibrational frequencies for Al$_{12}$Cu are 54.22-305.41 cm$^{-1}$, while for Al$_{12}$Zn are 49.59-321.09 cm$^{-1}$, respectively, denoting a narrow range of their vibrational spectra. The representative vibrational modes of the Al$_{12}$Cu cluster primarily correspond to anti-symmetric stretching,\cite{Guevara-Vela2024} as shown in Figure~\ref{figure2}.

The ionization energy (IP) and electron affinity (EA) are key physical parameters that indicate the electronic stability of clusters. The calculated values are presented in Table~\ref{table_1}, while their formulas and definitions can be found elsewhere.\cite{RODRIGUEZKESSLER2024122062,RODRIGUEZKESSLER2023116538,RODRIGUEZKESSLER2023121620,https://doi.org/10.1002/adts.202100283,D0CP04018E,D2CP05188E,D1CP00379H,RODRIGUEZKESSLER2020155897} The results show that the Al$_{12}$M (M=Cu,Zn) clusters have smaller IP values compared to the neutral Al$_{13}$ cluster, suggesting and increase in reactivity. Al$_{12}$Zn shows the smaller EA value, followed by Al$_{12}$Cu and Al$_{13}$. The chemical hardness ($\eta$) for Al$_{12}$M (M=Cu,Zn) clusters show similar values compared to the bare Al$_{13}$ clusters, while the chemical potential ($\mu$) also showed a small reduction for the doped clusters. Moreover, the structures of the clusters exhibited only one rotational axis, consistent with C$_1$ symmetry. The electronic states are $^2$A and $^1$A for the Al$_{12}$Cu and Al$_{12}$Zn clusters, respectively.

	\begin{table}[ht!]
	{
 \caption{\label{table_1}{The symmetry point group, electronic state, ionization energy, electron affinity, chemical hardness and chemical potential of Al$_{12}$M (M=Cu, Zn) and Al$_{13}$ clusters. The results are calculated at the PBE0 level in conjunction with the Def2-TZVP basis set. The energy is given in eV.}}
\small
\def\arraystretch{1.1}
\begin{tabular}{p{2.0cm}p{1.4cm}p{1.4cm}p{1.4cm}p{1.4cm}p{1.4cm}p{1.4cm}}
	{Cluster} &  Sym  & E$_{state}$	  &  IP     & EA    & $\eta$ & $\mu$    \\ \hline
  Al$_{12}$Cu$^{b}$ & C$_{1}$ & $^2$A  &  6.64  & 3.02  &  1.81  &  4.83  \\  
  Al$_{12}$Zn$^{a}$ & C$_1$ & $^1$A   &  6.66  & 2.83  &  1.91  &  4.74  \\
 \\
 Al$_{13}$ & C$_1$ & $^2$A   &  6.89  & 3.10  &  1.89  &  5.00  \\
  \hline
\end{tabular}

	}
\end{table}

\begin{figure}[h!]
\caption{\label{figura4}The total and partial density of states (TDOS, PDOS) for Al$_{12}$M (M=Cu, Zn) clusters. Vertical dashed line corresponds to the HOMO energy level.}
 \resizebox*{0.50\textwidth}{!}{\includegraphics{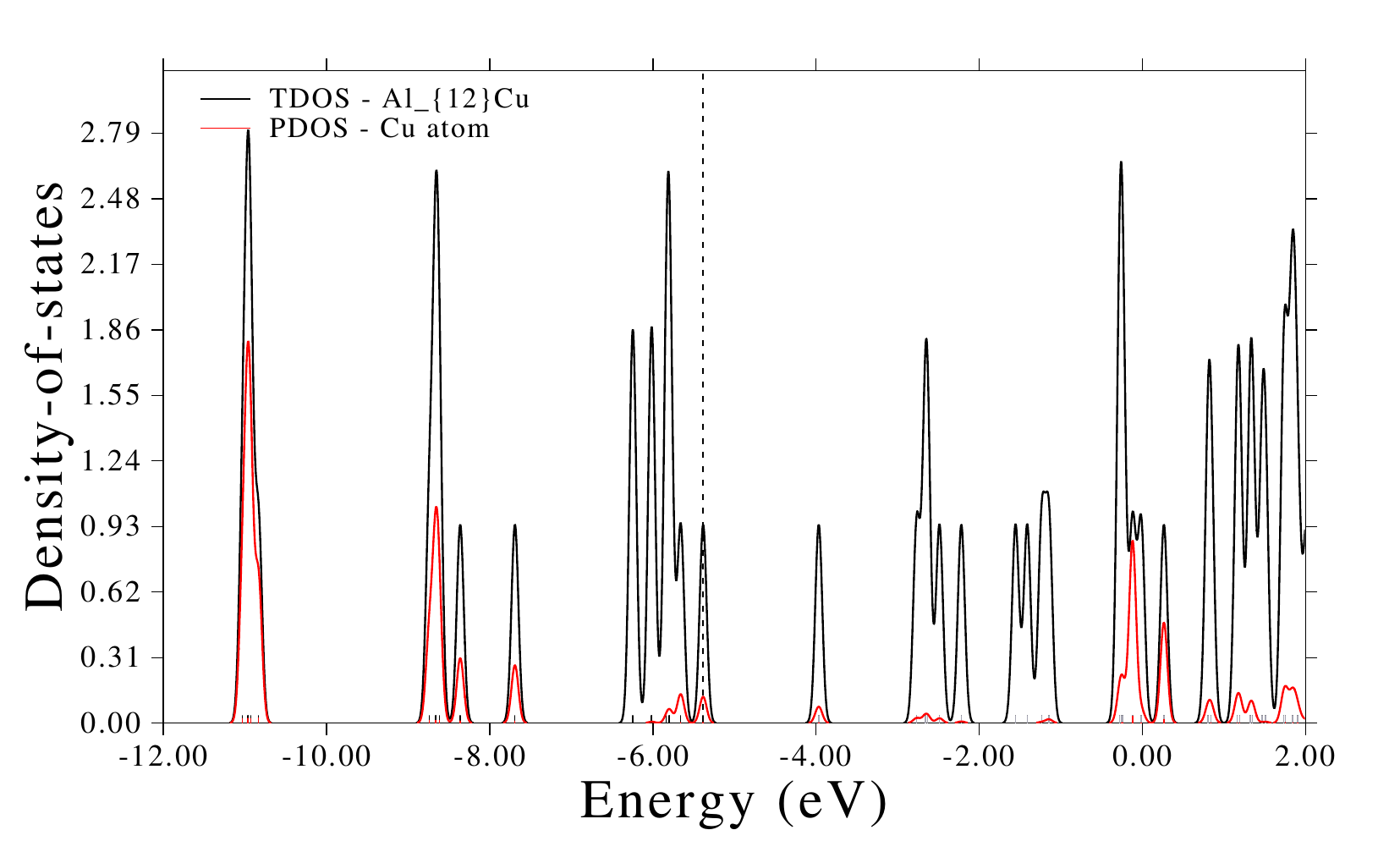}} 
\resizebox*{0.50\textwidth}{!}{\includegraphics{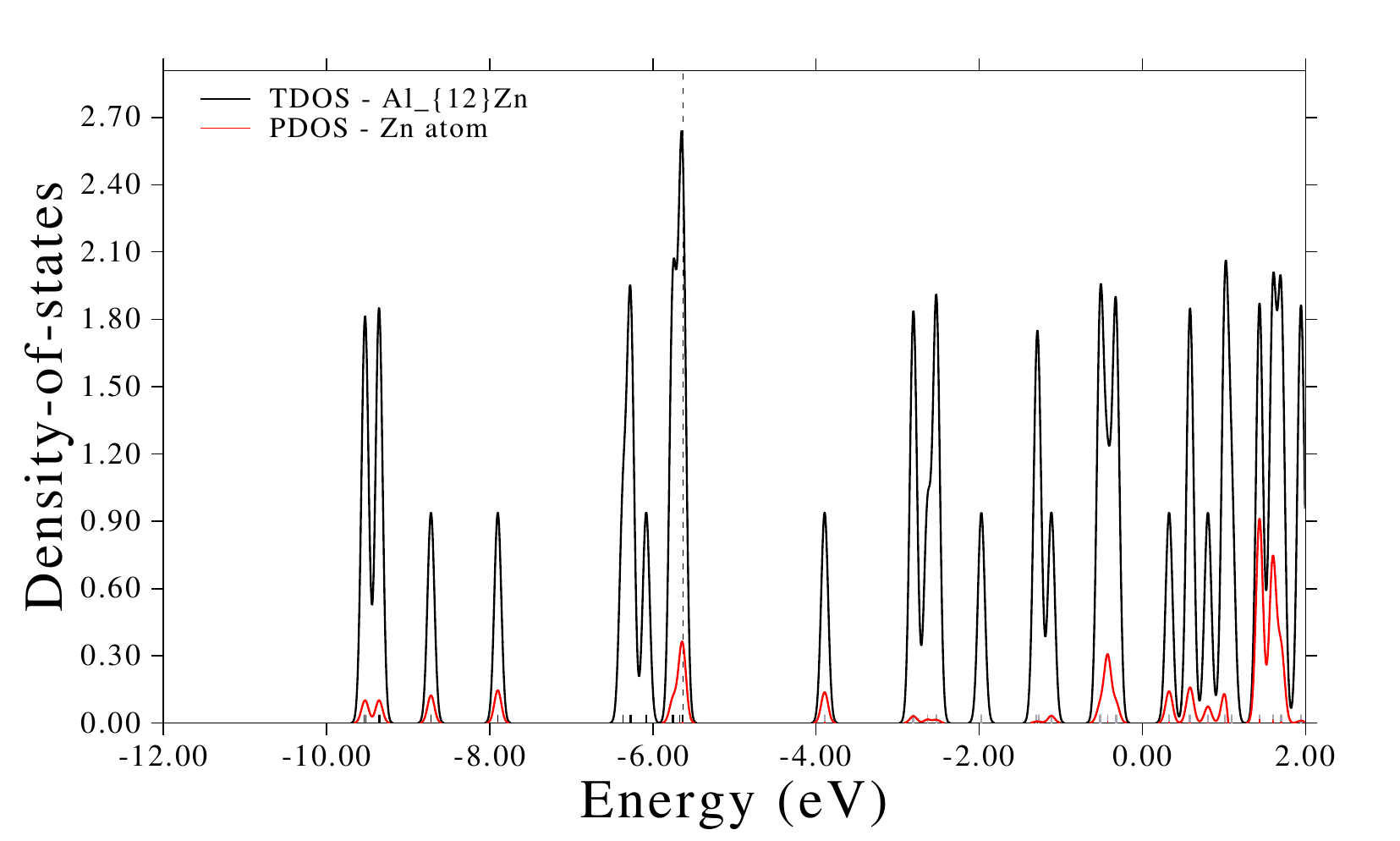}} 
\end{figure}

To get insight into the electronic properties of the clusters, the total density of states (TDOS) for Al$_{12}$M (M=Cu,Zn) clusters are depicted in Figure~\ref{figura4}. For M=Cu, the PDOS show deeper hybridization for the Cu atom, on the other hand for M=Zn, more states near the HOMO energy are found, suggesting that the former (Al$_{12}$Cu) is more stable. This preliminary results help for further exploring the reactivity of the clusters, with possible applications in catalysis.\cite{10.1063/1.4935566,D2CP05188E,RODRIGUEZKESSLER201820636,RODRIGUEZCARRERA2024122301,OLALDELOPEZ2024,doi:10.1021/acs.jpcc.8b09811}


\section*{Conclusions}
 In this preprint, the structural and electronic properties of Al$_{12}$M (Zn, Cu) clusters were investigated using density functional theory (DFT) calculations. The results showed that the clusters favor the low-spin states by considering three different functionals, the PBE, PBE0, and TPSSh respectively. The doped clusters showed lower ionization energy and electron affinity compared to the neutral Al$_{13}$ cluster. The density of states showed a higher degree of hybridization in Al$_{12}$Cu compared to Al$_{12}$Zn.


\section*{Acknowledgments}
P.L.R.-K. would like to thank the support of CIMAT Supercomputing Laboratories of Guanajuato and Puerto Interior. 

\nolinenumbers

\bibliography{mendeley}

\bibliographystyle{ieeetr}

\end{document}